\documentstyle[graphicx,editedvolume,numreferences,cite]{crckapb}
\newcommand{\be}{\begin{equation}}
\newcommand{\ee}{\end{equation}}
\newcommand{\ben}{\begin{eqnarray*}}
\newcommand{\een}{\end{eqnarray*}}
\newcommand{\bea}{\begin{eqnarray}}
\newcommand{\eea}{\end{eqnarray}}

\newcommand{\up}{\uparrow}
\newcommand{\dn}{\downarrow}
\newcommand{\beq}{\begin{equation}}
\newcommand{\eeq}{\end{equation}}

\begin{opening}
\title{Ab-initio Gutzwiller method: first application to
Plutonium}
\author{J.-P. JULIEN}
\institute{Laboratoire d'Etudes des Propri\'et\'es Electroniques
des Solides,\protect \\
CNRS, B.P. 166, 38042 Grenoble Cedex 9,
France}
\author{J. Bouchet}
\institute{CEA-DAM, DPTA,\protect \\
Bruy\`eres-le-Ch\^atel, France}
\end{opening}
\runningtitle{AB-INITIO GUTZWILLER METHOD...}

\begin{document}

\bigskip

\begin{abstract}
 Using a density matrix approach to
Gutzwiller method, we present a formalism to treat
\textit{ab-initio} multiband Tight-Binding Hamiltonians including
local Coulomb interaction in a solid, like, for e.g., the
degenerate Hubbard model. We first derive the main results of our
method: starting from the density matrix of the non-interacting
state, we build a multi-configurational variational wave function.
The probabilities of atomic configurations are the variational
parameters of the method. The kinetic energy contributions are
renormalized whereas the interaction contributions are exactly
calculated. A renormalization of effective on-site levels, in
contrast to the usual one-band Gutzwiller approach, is derived.
After minimization with respect to the variational parameters, the
approximate ground state is obtained, providing the equilibrium
properties of a material. Academic models will illustrate the  key
points of our approach. Finally, as this method is not restricted
to parametrized Tight-Binding Hamiltonians, it can be performed
from first principles level by the use of the so-called
"Linearized Muffin Tin Orbitals" technique. To avoid double
counting of the repulsion, one subtracts the average interaction,
already taken into account in this density functional theory
within local density approximation (DFT-LDA) based band structures
method and one adds an interaction part "a la Hubbard". Our method
can be seen as an improvement of the more popular LDA+$U$ method
as the density-density correlations are treated beyond a standard
mean field approach. First application to Plutonium will be
presented with peculiar attention to the equilibrium volume, and
investigations for other densities will be discussed.
\end{abstract}

\maketitle
\section{Introduction}
 Except for small molecules, it is impossible to solve many
electrons systems without imposing severe approximations. If the
configuration interaction approaches (CI) or Coupled Clusters
techniques \cite{FuldeBook} are applicable for molecules, their
generalization for solids is difficult. For materials with a
kinetic energy greater than the Coulomb interaction, calculations
based on the density functional theory (DFT), associated with the
local density approximation (LDA) \cite{Hohenberg64, Kohn65} give
satisfying qualitative and quantitative results to describe ground
state properties. These solids have weakly correlated electrons
presenting extended states, like $sp$ materials or covalent
solids. The application of this approximation to systems where the
wave functions are more localized ($d$ or $f$-states) as
transition metals oxides, heavy fermions, rare earths or actinides
is more questionable and can even lead to unphysical results : for
example, insulating FeO and CoO are predicted to be metalic by the
DFT-LDA. On another hand, theoretical "many body" approaches like
diagrammatic developments \cite{Mattuck}, slave bosons
\cite{Kotliar86}, decoupling of the equations of motion of Green
functions by projection techniques \cite{Mori}, and more recently
dynamical mean field theory (DMFT) in infinite dimension
\cite{Georges96}, treat in a much better way correlation effects
than the DFT-LDA does. However the price to be paid is an
oversimplification of the system, generally reducing the number of
involved orbitals and using parameterized Hamiltonians (like
Hubbard model) where the \textit{ab-initio} aspect of the DFT-LDA
is lost. Finally, these methods, contrary to the DFT-LDA, are
scarcely variational. Recently several attemps have been proposed
to couple these two points of view as in the LDA+$U$
\cite{Anisimov97b}, or LDA+DMFT \cite{Anisimov97a, Savrasov01}
approaches. In the same spirit, the approach we describe below,
tries to keep advantages on both sides: it is a variational method
which is multi-configurational, contrary to the DFT-LDA, but
without loosing the "adjustable parameters free" advantages of the
\textit{ab-initio} side. The next section is devoted to the
derivation of our formalism. It is then applied to known academic
cases to prove the reliability of our approach. The insertion of
this approach at the \textit{ab-initio} level is presented in
section 2.5. The nature of the electronic structure of Plutonium
being still under discussion, the application of our method, in
the last section, is in accordance with previous works and also
gives some new insights for this material.

\section{Method}
\subsection{Gutzwiller approach for the one-band Hubbard model}

Among numerous theoretical approaches, the Gutzwiller method
\cite{Gutzwiller63, Gutzwiller65} provides a transparent physical
interpretation in term of atomic configurations of a given site.
Originally it was applied to the one-band Hubbard model
Hamiltonian \cite{Hubbard64}:

\begin{equation}
\label{hub} H=H_{kin}+H_{int}
\end{equation}

\noindent with

\begin{equation}
H_{kin}=\sum_{i\neq j,\sigma} t_{ij} c_{i\sigma}^\dagger
c_{j\sigma}
\end{equation}
\noindent and

\begin{equation}
H_{int}=U\sum_i n_{i\uparrow}n_{i\downarrow}
\end{equation}

\noindent which contains a kinetic part $H_{kin}$ with a hopping
integral $t_{ij}$ from site $j$ to site $i$, and an interaction
part with a local Coulomb repulsion $U$ for electrons on the same
site. $c_{i\sigma}^\dagger$ (respectively $c_{i\sigma}$) is the
creation (respectively destruction) operator of an electron at
site $i$ with up or down spin $\sigma$.
$n_{i\sigma}=c_{i\sigma}^\dagger c_{i\sigma}$ measures the number
(0 or 1) of electron at site $i$ with spin $\sigma$. This
Hamiltonian contains the key ingredients for correlated up and
down spin electrons on a lattice: the competition between
delocalization of electrons by hoppings and their localization by
the interaction. It is one of the most used models to study
electronic correlations in solids (for a review see Ref.
\cite{bookGebhard}).

In the absence of the interaction $U$, the ground state is that of
uncorrelated electrons $|\Psi_0\rangle$ and has the form of a
Slater determinant. As $U$ is turned on, the weight of doubly
occupied sites must be reduced because they cost an additional
energy $U$ per site. Accordingly, the trial Gutzwiller wave
function (GWF) $|\Psi_{G}\rangle$ is built from the Hartree-like
uncorraleted wave function (HWF) $|\Psi_0\rangle$,

\begin{equation}
|\Psi_{G}\rangle = g^D|\Psi_0\rangle
\end{equation}

The role of $g^D$ is to reduce the weight of configurations (i.e.
a way of spreading $N$ electrons over the lattice) with doubly
occupied sites, where $D=\sum_i n_{i\up} n_{i\dn}$ measures the
number of double occupations and $g$ ($<1$) is a variational
parameter. In fact, this method corrects the mean field (Hartree)
approach for which up and down spin electrons are independent,
and, some how, overestimates configurations with double occupied
sites. Using the Rayleigh-Ritz principle, this parameter is
determined by minimization of the energy in the Gutzwiller state
$|\Psi_{G}\rangle$, giving an upper bound to the true unknown
ground state energy of $H$. Note that to enable this calculation
to be tractable, it is necessary to use the Gutzwiller's
approximation which assumes that all configurations in the HWF
have the same weight. Details of the derivation can be found in
the article of Vollhardt \cite{Vollhardt84}.

Nozi\`eres \cite{Nozieres} proposed an alternative way, showing
that the Gutzwiller approach is equivalent to renormalize the
density matrix in the GWF which can be reformulated as:

\beq \label{rho} \rho_G=T^{\dagger}\rho_0T \eeq

\noindent The density matrices
$\rho_G=|\psi_{G}\rangle\langle\psi_G|$ and
$\rho_0=|\psi_{0}\rangle\langle\psi_0|$ are projectors on the GWF
and HWF respectively. $T$ is an operator, diagonal in the
configuration basis, $T=\prod_i T_i$ and $T_i$ is a diagonal
operator acting on site $i$ \beq \label{T}
T_i|L_i,L'\rangle=\sqrt{\frac{p(L_i)}{p_0(L_i)}}|L_i,L'\rangle
\eeq \noindent $L_i$ is an atomic configuration of the site $i$,
with probability $p(L_i)$  in the GWF and $p_0(L_i)$ in the HWF
respectively, whereas $L'$ is a configuration of the remaining
sites of the lattice. Note that this prescription does not change
the phase of the wave function as the eigenvalues of the operators
$T_i$ are real. The correlations are local, and the configuration
probabilities for different sites are independent.

\noindent The expectation value of Hamiltonian (\ref{hub}) is
given by

\begin{equation}
\label{}
\langle H\rangle_G=Tr(\rho_G H)
\end{equation}

\begin{figure}[h]
\begin{center}
\includegraphics[width=8cm]{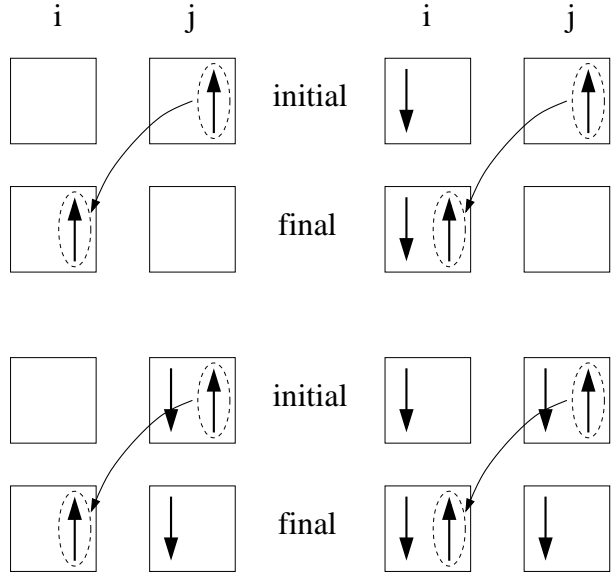}
\caption{\label{config} Initial and final configurations
contributing to the mean value $\langle c_{i\sigma}^\dagger
c_{j\sigma}\rangle_G$.}
\end{center}
\end{figure}

\noindent The mean value of one-site operators (interaction $U$)
is exactly calculated with the double occupancy probability
$d_i=\langle n_{i\up} n_{i\dn} \rangle_G$. $d_i$ is the new
variational parameter replacing $g$. From expressions (\ref{rho})
and (\ref{T}), the two-sites operators contributions of the
kinetic energy can be written as

\begin{equation}
\langle c_{i\sigma}^\dagger c_{j\sigma}\rangle_G=Tr(\rho_G
c_{i\sigma}^\dagger c_{j\sigma})=\langle c_{i\sigma}^\dagger
c_{j\sigma}\rangle_0 \sum_{L_{-\sigma}}\sqrt{\frac{p(L'_\sigma ,
L_{-\sigma})}{p_0(L'_\sigma)}}\sqrt{\frac{p(L_\sigma ,
L_{-\sigma})}{p_0(L_\sigma)}} \end{equation}

\noindent where $L'_{\sigma}$ and $L_{\sigma}$ are the only two
configurations of  spin $\sigma$ at sites $i$ and $j$ that give
non-zero matrix element to the operator in the brackets as
illustrated on Fig.\ref{config}. The summation is performed over
the configurations of opposite spin $L_{-\sigma}$. Their
corresponding probabilities are pictured on Table \ref{proba} for
an homogeneous state (for any site $i$, $\langle n_{i\sigma}
\rangle=n$ and $\langle n_{i\uparrow}n_{i\downarrow}\rangle=d$).
The probabilities $p_0$ in the HWF depend only on the number of
electrons, whereas the $p$ in the GWF also depend on $d_i$.

\begin{table}
\begin{center}
\centering
\begin{tabular}{|c|c|c|}
\hline $L_i$ & $p_0(L_i)$ & $p(L_i)$ \\ \hline $\emptyset$ &
$(1-n)^2$ & $1-2n+d$\\ \hline $\uparrow$ & $n(1-n)$ & $n-d$\\
\hline $\downarrow$ & $n(1-n)$ & $n-d$\\
\hline $\uparrow\downarrow$ & $n^2$ & d \\ \hline
\end{tabular}
\caption{\label{proba}Different possible configurations of one
site and the corresponding probabilities in the HWF ($p_0$) and
GWF ($p$).}
\end{center}
\end{table}

After some elementary algebra, one can show that the Gutzwiller
mean value can be factorised:

\begin{equation}
\langle c_{i\sigma}^\dagger
c_{j\sigma}\rangle_G=\sqrt{q_{i\sigma}}\langle c_{i\sigma}^\dagger
c_{j\sigma}\rangle_0\sqrt{q_{j\sigma}}
\end{equation}
\noindent where these renormalization factors $q_{i\sigma}$ are
local and can be expressed as:

\begin{equation}
\sqrt{q_{i\sigma}}=\frac{\left(\sqrt{1-n_{i\sigma}-n_{i-\sigma}+d_i}+\sqrt{d_i}
\right)\sqrt{n_{i-\sigma}-d_i}}{\sqrt{n_{i\sigma}(1-n_{i\sigma})}}
\end{equation}
\noindent where $n_{i\sigma}$ is a shorthand for $\langle
n_{i\sigma} \rangle$, the average number of electron on the
considered "orbital-spin" in the HWF, which could be site and/or
spin independent if the state is homogeneous and/or paramagnetic
(it is the case we consider here for pedagogy, dropping indices
$i\sigma$). The kinetic energy $\epsilon^0_{kin}$ of the
non-interacting electrons state is renormalized by a factor $q$
which is smaller than one in the correlated state, and equal to
one in the HWF. Then, we minimize the variational energy

\begin{equation}\label{}
E(d)=\langle H \rangle_G=q\epsilon^0_{kin}+Ud
\end{equation}
with respect to $d$. In the case of half filling ($n=1/2$), if the
repulsion $U$ exceeds a critical value $U_c=8 \epsilon^0_{kin}$,
$q$ is equal to zero, leading to an infinite quasiparticle mass
with a Mott-Hubbard Metal-Insulator transition which is, in this
context, often referred to as "the Brinkmann-Rice transition"
\cite{BrinkmannRice70}, as these authors first applied the
Gutzwiller approximation to the Metal-Insulator transition.
Application of this "one orbital per site" formalism for
inhomogeneous states is possible because all involved quantities
are local. An example can be found in \cite{Mayou88} for model
$CuO_2$ planes, in connection with the electronic structure of
High $T_C$ superconductors.

\subsection{Inequivalent sites: renormalization of levels}
When sites are inequivalent, or if orbitals belong to different
symmetries as in a multiorbital $spdf$ basis case of further
sections, it is necessary to add to the Hamiltonian an on-site
energy term
\beq
H_{on-site}=\sum_{i\sigma}\epsilon^0_{i\sigma}n_{i\sigma} \eeq

\noindent Hence this enlarged Hubbard Hamiltonian can be written
as

\beq \label{hub+onsite}
 H=\sum_{i\neq j,\sigma} t_{ij}
c_{i\sigma}^\dagger
c_{j\sigma}+\sum_{i\sigma}\epsilon^0_{i\sigma}n_{i\sigma}+U\sum_i
n_{i\up} n_{i\dn}
\eeq

\noindent In that case, the starting HWF, directly obtained from
the non-interacting part of the Hamiltonian, is not automatically
the best choice, giving the optimal GWF, i.e. having the lowest
energy.
For example, if we look for the ground state of Hamiltonian
(\ref{hub+onsite}) in the Hartree-Fock (HF) self-consistent field
formalism, it is necessary to vary the orbital occupations.
Practically, it can be achieved by replacing this Hamiltonian, by
an effective Hamiltonian $H_{eff}$ of independent particles with
renormalized on-site energies $\epsilon_{i\sigma}$:

\beq \label{Heff} H_{eff}=\sum_{i\neq j,\sigma} t_{ij}
c_{i\sigma}^\dagger
c_{j\sigma}+\sum_{i\sigma}\epsilon_{i\sigma}n_{i\sigma}(+C)
\eeq

The HWF we are looking for, is an \textit{approximate} ground
state of the \textit{true} many-body Hamiltonian
(\ref{hub+onsite}) and is the \textit{exact} ground state of
\textit{effective} Hamiltonian (\ref{Heff}). The additive constant
$C$ accounts for double counting energy reference, so that the
ground state energies are the same for both Hamiltonians:

\beq \label{heff=h}
\langle H_{eff} \rangle=\langle H \rangle
\eeq

\noindent The effective Hamiltonian depends on pararemeters
$\epsilon_{i\sigma}$. The optimal choice can be obtained by
minimizing the ground state energy of $H_{eff}$ with respect to
these parameters. With the help of Hellmann-Feyman theorem, one
can easily see that the derivative of the kinetic energy is

\beq \label{derivHkin}
\frac{\partial\langle
H_{kin}\rangle}{\partial\epsilon_{i\sigma}}= -\sum_{j\neq
i,\sigma}\epsilon_{j\sigma}\frac{\partial\langle
n_{j\sigma}\rangle}{\partial\epsilon_{i\sigma}} \eeq

\noindent On another hand, differentiation of equality
(\ref{heff=h}) associated with expression (\ref{derivHkin}) and
the mean field approximation  $\langle n_{i\up}
n_{i\dn}\rangle\approx\langle n_{i\up}\rangle\langle
n_{i\dn}\rangle $ enables to retrieve the well-known formula for
the on-site energies

\beq \label{renoHF}
\epsilon_{i\sigma}=\epsilon^0_{i\sigma}+U\langle
n_{i-\sigma}\rangle
\eeq

\noindent and the constant C is simply $-U\sum_i\langle
n_{i\up}\rangle\langle n_{i\dn}\rangle $.

In the Gutwiller approach, the same argument about the variation
of orbital occupation, i.e. flexibility on the HWF
$|\Psi_0\rangle$, is true. It is necessary to find a way to vary
this Slater determinant from which the GWF $|\Psi_G\rangle$ is
generated, so that the Gutzwiller ground-state energy is minimum.
Clearly one has to find an equivalent of formula (\ref{renoHF}) in
the Gutzwiller context, which has never been established, to our
knowledge. The average value of Hamiltonian (\ref{hub+onsite}) on
a GWF is given by:

\beq \label{average-hub+onsite} \langle
\Psi_G|H|\Psi_G\rangle=\sum_{ij\sigma}t_{ij}\sqrt{q_{i\sigma}}\langle
c_{i\sigma}^\dagger c_{j\sigma}\rangle_0\sqrt{q_{j\sigma}}+U\sum_i
d_i+\sum_{i\alpha\sigma}\epsilon^0_{i\sigma}\langle
n_{i\sigma}\rangle_0 \eeq

Following the same footing of previous HF self-consistent field
approach, one has to find an effective Hamiltonian $H_{eff}$ of
independent particles having $|\Psi_0\rangle$ as \textit{exact}
ground state. This state $|\Psi_0\rangle$ generates the GWF
$|\Psi_G\rangle$ which is an \textit{approximate} ground state of
the true interacting Hamiltonian (\ref{hub+onsite}). In analogy
with (\ref{heff=h}), the condition

\beq \label{heff=h2}
\langle \Psi_0|H_{eff}|\Psi_0\rangle=\langle
\Psi_G|H|\Psi_G\rangle
\eeq

\noindent leads to the expression for the searched $H_{eff}$:

\beq \label{Heff2}
H_{eff}=\sum_{i\neq j,\sigma} \tilde{t_{ij}}
c_{i\sigma}^\dagger
c_{j\sigma}+\sum_{i\sigma}\epsilon_{i\sigma}n_{i\sigma}+C^{\prime}
\eeq

\noindent with effective but \textit{fixed} renormalized hoppings
$\tilde{t_{ij}}=\sqrt{q_{i\sigma}} t_{ij} \sqrt{q_{j\sigma}}$ and
having effective on-site energies $\epsilon_{i\sigma}$ which have
still to be determined. Hellmann-Feynman theorem applied to
$H_{eff}$ provides again an expression similar to
(\ref{derivHkin}), but with effective hoppings. Taking into
account the dependence of the ${q_{i\sigma}}$'s through
${n_{i\sigma}}$ in differentiating (\ref{average-hub+onsite}) and
(\ref{heff=h2}) with respect to the parameters
$\epsilon_{i\sigma}$, after some calculations, one obtains the
equivalent expression of (\ref{renoHF}) in the Gutzwiller context:

\beq \label{renoG}
\epsilon_{i\sigma}=\epsilon^0_{i\sigma}+2e_{i\sigma}\frac{\partial
ln(\sqrt{q_{i\sigma}})}{\partial
n_{i\sigma}} \eeq

Here $e_{i\sigma}$ is the partial kinetic energy of orbital-spin
$i \sigma$, it is given by

\beq \label{partial-kin} e_{i\sigma}=\sum_{j\sigma}\tilde{t_{ij}}
\langle c_{i\sigma}^\dagger
c_{j\sigma}\rangle_0=\int_{-\infty}^{\rm E_F}
E\tilde{N_{i\sigma}}(E)dE-\epsilon_{i\sigma}\langle
n_{i\sigma}\rangle_0 \eeq

with $\tilde{N_{i\sigma}}$ the $i \sigma$-projected density of
states (DOS) for Hamiltonian $H_{eff}$. The remaining constant
$C^{\prime}$ that ensures (\ref{heff=h2}) explicitly reads

\begin{equation}
C^{\prime}=U\sum_i d_i-\sum_{i\sigma} 2e_{i\sigma}\frac{\partial
ln(\sqrt{q_{i\sigma}})}{\partial n_{i\sigma}}
\end{equation}

To solve the full problem of finding an approximate ground state
to Hamiltonian (\ref{hub+onsite}), one is faced to a
self-consistent loop which can be proceeded in two steps. First
one can get the occupations $\langle n_{i\sigma}\rangle_0$ from a
HWF, and a set of 'bare' $\epsilon^{0}_{i\sigma}$ levels. Then one
obtains  a set of configuration parameters, the probabilities of
double occupation, $d_i $ by minimizing (\ref{average-hub+onsite})
with respect to these probabilities. Afterwards the on-site levels
are renormalized according to (\ref{renoG}) and the next loop
starts again for the new effective Hamiltonian $H_{eff}$ till
convergence is achieved.

\begin{figure}[h]
\begin{center}
\rotatebox{270}{\includegraphics[width=8cm]{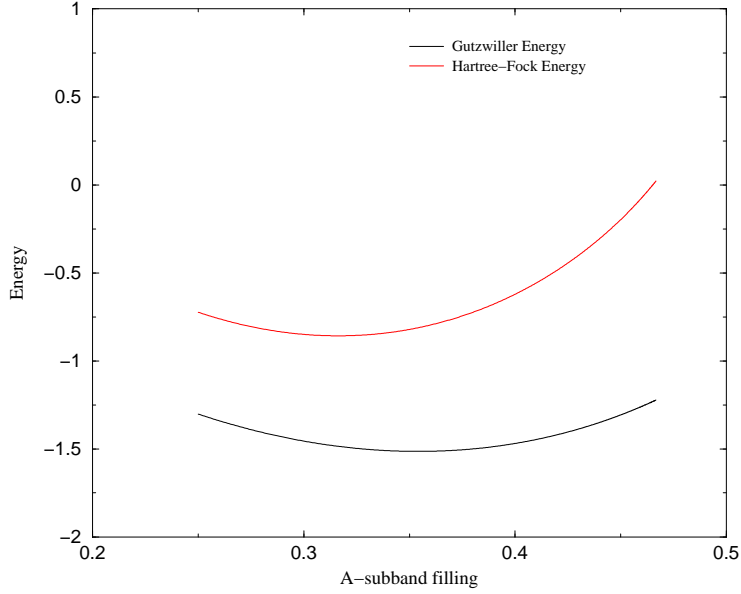}}
\caption{\label{nrj-chain}Total energy of the alternate chain
versus A-subband filling. Upper curve: Hartree-Fock result, Lower
curve: Gutzwiller result. The 2 minima are clearly different }
\end{center}
\end{figure}

As illustration of the importance of the renormalization of
levels, we studied the case of an alternate Hubbard chain (often
called Ionic Hubbard model). This model contains two kinds of
alternating atoms A and B on an infinite chain, having on-site
energies $\epsilon^{0}_A$ and $\epsilon^{0}_B$ respectively,
coupled by a hopping integral $t$. The same local Coulomb
repulsion U acts on each site.
For a given total number of electrons, one can fix a repartition
of electrons among sites A and B, and compute the energy of the
ground state: first within the HF mean field approximation, and
secondly, within the Gutzwiller approach. Browsing the electronic
occupation of A-site, by adjunction of Lagrange multiplier to fix
it to a given value, one looks for the lowest energy state. The
corresponding ground state energies as function of the A-filling
are presented on Fig. \ref{nrj-chain}. First of all, the lowest
Hartree state could be more efficiently directly found, after some
self-consistent loops, via the on-site renormalization of levels
of equation (\ref{renoHF}) in the HF context, as explained in
precedent paragraph. By inspection of the curve, it is also
obvious that this lowest Hartree does not generate the lowest
Gutzwiller state. It is necessary to browse among different
A-fillings to find the best Gutzwiller ground state. If this
browsing procedure is still tractable for simple models, as we did
in Ref. \cite{Mayou88}, its generalization to multiorbitals cases
would be practically impossible. It is the main advantage of
formula (\ref{renoG}) to avoid this cumbersome search for
optimized levels and to provide a systematic way of finding them,
similar to (\ref{renoHF}), leading to the best (i.e. lowest)
Gutzwiller ground state.

\subsection{Generalization to the degenerate Hubbard Hamiltonian}

Now we generalize this density matrix formalism \cite{Julien99,
Julien05} for the degenerate Hubbard Hamiltonian which, with usual
notations, reads:

\begin{equation}
\label{hubg}
H=\sum_{i\neq j
\alpha\beta\sigma} t_{i\alpha,j\beta}c_{i\alpha\sigma}^\dagger
c_{j\beta\sigma}+H_{\mathrm{int}}
\end{equation}

with the model interaction

\begin{equation}\label{Hint2}
H_{\mathrm{int}}=\frac{1}{2}\sum_{i,\alpha\sigma\neq\beta\sigma'}
U_{\alpha\sigma\beta\sigma'}n_{i\alpha\sigma}n_{i\beta\sigma'}
\end{equation}

where $\alpha$, $\beta$ and $\sigma$, $\sigma'$ are orbitals and
spins index respectively, necessary to account for orbital
degeneracy. The case $\alpha\sigma=\beta\sigma'$ is excluded from
the interaction because of Pauli principle.  We neglect any spin
flip term in the interaction for simplicity. They could be in
principle taken into account in our approach, as it is done in a
different work by B\"{u}nemann et al \cite{Bunemann98b}. However
this procedure would involve a diagonalization of atomic part of
Hamiltonian, that complicates the presentation of  our approach
without bringing any new physical ingredients.

As in the one-band case, we define a Gutzwiller renormalized
density matrix with the operator $T$ given by eqs (\ref{rho}) and
(\ref{T}). The main difference being the greater number of atomic
configurations, equal to $2^{2N}$, with $N$ the orbital
degeneracy. For a given site we have now probabilities for double,
triple, etc... multiple occupancy, which are the new variational
parameters generalizing the role of $d$. Of course, the number of
independent variational probabilities is smaller than the number
of configurations, as different configurations could have the same
probabilities for symmetries reasons. For example, in a
paramagnetic case, a configuration and its spin reverse are
equivalent leading to the same probability. Moreover, the
probabilities are not independent of each other as the sum over
all probabilities have to be equal to 1, and we have also to
conserve the average electronic occupation of given orbital-spin
$\langle n_{i\alpha\sigma}\rangle$. These constraints could be
either directly included in the expressions of empty and single
occupied configurations probabilities, or treated by adjunction of
Lagrange multipliers as in the slaves bosons approach
\cite{Kotliar86}. This last formulation has the advantage of
giving more symmetric expressions. Using the expression (\ref{T})
of $T_i$ operators, we can directly obtained the factorized form
of the kinetic energy terms:

\begin{equation}
\langle c_{i\alpha\sigma}^\dagger
c_{j\beta\sigma}\rangle_G=\sqrt{q_{i\alpha\sigma}}\langle
c_{i\alpha\sigma}^\dagger
c_{j\beta\sigma}\rangle_0\sqrt{q_{j\beta\sigma}}
\end{equation}

\noindent where the $q$-factors reduce the kinetic energy and are
expressed as functions of the variational parameters and the
number of electrons according to:

\begin{equation}
\label{qgeneral}
\sqrt{q_{i\alpha\sigma}}=\frac{1}{\sqrt{n_{i\alpha\sigma}(1-n_{i\alpha\sigma})}}
\sum_{L'_i}\sqrt{p(i\alpha\sigma:unocc,L'_i)p(i\alpha\sigma:occ,L'_i)}
\end{equation}

\noindent Here $p(i\alpha\sigma:occ,L'_i)$  (respectively
$p(i\alpha\sigma:unocc,L'_i)$) represents the probability of the
atomic configuration of site $i$, where the orbital $\alpha$ with
spin $\sigma$ is occupied (resp. unoccupied) and where $L'_i$ is a
configuration of the remaining orbitals of this site. This result
is similar to the expression obtained by B\"unemann \textit{et
al.}\cite{BGW-JPhConMatter97}, but it is obtained more directly by
the density matrix renormalization (\ref{rho}). To obtain the
expression of the $q_{i\alpha\sigma}$ factors, an additional
approximation to the density matrix of the uncorrelated state was
necessary. This approximation can be viewed as the multiband
generalization of the Gutzwiller approximation, exact in infinite
dimension \cite{MetznerVollhardt87}

\begin{equation}
\langle L L''|\rho_0|L'L''\rangle\approx
p_0(L'')\sum_{L''}\langle L L''|\rho_0|L'L''\rangle
\end{equation}

\noindent Where we have replaced an off-diagonal element of the
density by its average value over the configurations $L''$. $L$
and $L'$ are configurations of one or two sites, involved in the
calculation of interaction or kinetic term and $L''$ is the
configuration of remaining sites.  This approximation allows to
perform calculations, and however preserves sum rules of the
density matrix. Similarly, the interaction between an electron at
site $i$ on orbital $\alpha$ with spin $\sigma$ and an electron on
orbital $\beta$ with spin $\sigma'$ involves a term

\begin{equation}
\label{mean-interaction} \langle
n_{i\alpha\sigma}n_{i\beta\sigma'}
\rangle=\sum_{L'_i}p(i\alpha\sigma:occ,i\beta\sigma':occ,L'_i)
\end{equation}

\noindent where $L'_i$ is a configuration of the remaining
spin-orbitals of this site, other than $\alpha\sigma$ and
$\beta\sigma'$.

As illustration, we studied the academic case of paramagnetic
state for doubly degenerate bands like, for instance,
$e_g$-symmetry $d$-orbitals in cubic or octahedral environment.
Hybridization among these degenerate orbitals is supposed to
produce a kinetic energy $\epsilon^0_{kin}$ in the uncorrelated
state. We take a
model interaction 
where the general expression (\ref{Hint2}) reduces to:

\begin{eqnarray}
&H_{\mathrm{int}}=U \sum_{i\alpha\sigma}
n_{i\alpha\sigma}n_{i\alpha -\sigma}+U^{\prime} \sum_{i \alpha
\neq \beta \sigma}n_{i\alpha\sigma}n_{i\beta -\sigma}\nonumber\\
&+(U^{\prime}-J) \sum_{i\alpha \neq \beta
\sigma}n_{i\alpha\sigma}n_{i\beta \sigma} \label{Hint2simplified}
\end{eqnarray}

with two independent parameters $U$ and $U^{\prime}$ as the
relation $U-U^{\prime}=2J$ stands \cite{Sugano}. The interaction
between electrons of same spin is reduced by the exchange integral
$J$, which is essential to reproduce first Hund's law of maximum
spin. The application of the above prescription directly leads to
the variational energy:

\begin{equation}\label{Eg2band}
E_G=2q\epsilon^0_{kin}+2Ud_0+2U^{\prime}
d_1+2(U^{\prime}-J)d_2+2(U+2U^{\prime}-J)(2t+f)
\end{equation}

 \noindent
$f$ and $t$ are the quadruple and triple occupancy respectively,
whereas there are three possibilities of double occupancies :
$d_0$ (same orbital, different spin), $d_1$ (different orbital,
different spin)and $d_2$ (different orbital, same spin). Some of
the corresponding configurations with multiple occupancy are
pictured on Table \ref{proba2}, followed by their probability and
their interaction energy. This expression is identical to the
result obtained by different authors using Gutzwiller-type wave
function\cite{Bunemann97c, Okabe}, or multiband slave-boson
approach\cite{Hasegawa97b} which is a multiorbital generalization
of Ref.\cite{Kotliar86}. It is to be stressed the very physical
"transparent" approach with the density matrix formalism, leading
to simple expressions. Also, there is no approximation about less
favorable configurations, discarded from the beginning as in Ref.
\cite{Lu}. For a given electronic filling, we use a Newton-Raphson
procedure to minimize $E_G$ with respect to $d_0$, $d_1$, $d_2$,
$t$ and $f$. We again choose a half filled case, and we scale all
contributions in term of the kinetic energy.


\begin{table}

\hspace{5cm}\begin{tabular}{|p{0.4cm}|p{0.4cm}|} \hline
$\uparrow\downarrow$  &\\
\hline
\end{tabular}
\hspace{1cm} $d_0$\hspace{1cm} $U$ \vspace{0.5cm}

\hspace{5cm}\begin{tabular}{|p{0.4cm}|p{0.4cm}|} \hline
$\uparrow$ &  $\downarrow$ \\
\hline
\end{tabular}
\hspace{1cm} $d_1$\hspace{1cm} $U'$ \vspace{0.5cm}

\hspace{5cm}\begin{tabular}{|p{0.4cm}|p{0.4cm}|} \hline
$\uparrow$ &  $\uparrow$ \\
\hline
\end{tabular}
\hspace{1cm} $d_2$\hspace{1cm} $U'-J$ \vspace{0.5cm}

\hspace{5cm}\begin{tabular}{|p{0.4cm}|p{0.4cm}|} \hline
$\uparrow\downarrow$ &  $\downarrow$ \\
\hline
\end{tabular}
\hspace{1cm} $t$\hspace{1cm} $U+2U'-J$ \vspace{0.5cm}

\hspace{5cm}\begin{tabular}{|p{0.4cm}|p{0.4cm}|} \hline
$\uparrow\downarrow$ &  $\uparrow\downarrow$ \\
\hline
\end{tabular}
\hspace{1cm} $f$\hspace{1cm} $2U+2(2U'-J)$
\caption{\label{proba2}}
\end{table}

\begin{figure}[h]
\begin{center}
\rotatebox{270}{\includegraphics[width=8cm]{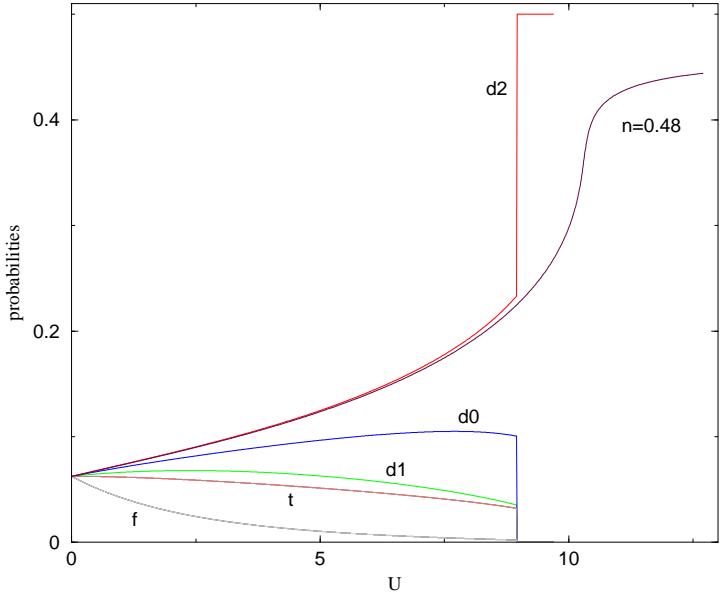}}
\caption{\label{result2bdes}For the half filling case all the
probabilities are equal for $U=0$. For $U_c \sim 9$ we have a
transition from the metallic to the insulator state which is of
first order at half filling and second order for other
concentration as seen for $n=0.48$. }
\end{center}
\end{figure}

On Fig. \ref{result2bdes} we plot the probabilities of different
configurations versus the direct Coulomb interaction $U$. It can
be seen that the system undergoes a metal-insulator transition for
a sufficiently high value of $U$, close to 9. It is easy to
perform the same kind of calculation in the case of triply
degenerate orbitals relevant for $d$-orbitals with the symmetry
$t_{2g}$, (for instance Ti $t_{2g}$ in LaTiO$_3$), $f$-orbitals
with the symmetry $T_1$ or $T_2$ in cubic or octahedral
environment (rare-earth element or actinides) or $p$-orbitals like
in fullerene C$_{60}$.

\subsection{Different orbital symmetries: fully hybridized Hamiltonian}

As our aim is to describe realistic materials from the
\textit{ab-initio} level, it is necessary to have a full set of
$spd$ and possibly $f$ (for actinides or rare earths) basis. The
system can be described by the following Hamiltonian which is the
sum of a kinetic term, local Coulomb repulsions and an on-site
contribution accounting of difference of site (and/or symmetries)

\begin{eqnarray}
\label{Hgeneral}
&H=\sum_{i\neq j \alpha \beta \sigma}t_{i \alpha,j \beta}c_{i \alpha \sigma}^\dagger c_{j \beta \sigma} \nonumber\\
&+\frac{1}{2}\sum_{i, \alpha \sigma \neq \beta \sigma'}U_{\alpha \sigma \beta \sigma'}n_{i \alpha \sigma}n_{i \beta \sigma'}\nonumber\\
&+\sum_{i \alpha \sigma}\epsilon^0_{i \alpha \sigma}n_{i \alpha
\sigma}
\label{Htotal}
\end{eqnarray}

The kinetic term is responsible of the hybridization of orbitals
with different $\ell$-symmetries on neignboring sites among each
others. Moreover,we assume that the interaction part of the
Hamiltonian $H_{int}$ only concerns one subset of correlated
orbitals (say  $f$). All atomic configurations $\Gamma$ of this
subset on each site $i$ will be considered. Note that $H$ can be
seen as a multiband hybrid between the Hubbard Hamiltonian and the
periodic Anderson Hamiltonian. It contains hybridization of
localized interacting $f$ orbitals among each others (Hubbard) but
also with extended $spd$ states (Anderson). Using the results of
previous sections, the variational energy in the Gutzwiller state
reads

\begin{eqnarray}
&E_G=\sum_{i \neq j\alpha \beta \sigma}\sqrt{q_{i \alpha
\sigma}}t_{i \alpha,j \beta}\langle c_{i \alpha \sigma}^\dagger
c_{j \beta \sigma}\rangle_0\sqrt{q_{j \beta
\sigma}}\nonumber\\
&+\sum_{i,\Gamma} U_{\Gamma} p_i(\Gamma)+\sum_{i\alpha
\sigma}\epsilon^0_{i \alpha \sigma}\langle n_{i \alpha
\sigma}\rangle_0
\label{Egeneral}
\end{eqnarray}

\noindent In this expression the $q$-factors of site $i$ are
functions, through (\ref{qgeneral}), of the probabilities
$p_i(\Gamma)$ of the atomic configurations $\Gamma$ of
$f$-orbitals at the same site. They are equal to 1 if orbital
$\alpha$ or $\beta$ does not belong to this subset, i.e. for
extended states. $U_{\Gamma}$ is a proper combination of Coulomb
direct and exchange contributions $U_{\alpha \sigma \beta
\sigma'}$, accounting for the interaction energy which arises as
prefactors of expressions (\ref{mean-interaction}), and which can
be seen for e.g. on simplified case of (\ref{Eg2band}). As in our
previous simpler models, the probabilities $p_i(\Gamma)$ are the
variational parameters and one has to minimize $E_G$ with respect
to each of them (and at each inequivalent site) according to

\begin{eqnarray}
0=\frac{\partial E_G}{\partial p_i(\Gamma)}= \sum_{\alpha j\beta
\sigma}\frac{\partial\sqrt{q_{i\alpha\sigma}}}{\partial
p_i(\Gamma)}t_{i \alpha,j \beta}\langle c_{i \alpha
\sigma}^\dagger c_{j \beta \sigma}\rangle_0\sqrt{q_{j \beta
\sigma}}\nonumber\\
+\sum_{j\beta \sigma}\sqrt{q_{j \beta \sigma}}t_{j \beta,i \alpha}
\langle c_{j \beta \sigma}^\dagger c_{i \alpha \sigma}\rangle_0
\frac{\partial\sqrt{q_{i\alpha\sigma}}}{\partial p_i(\Gamma)}+
U_{\Gamma} \label{derivp}
\end{eqnarray}
To avoid the cumbersome calculations of all $\langle
c_{i\alpha\sigma}^\dagger c_{j\beta\sigma} \rangle_0$ cross-terms
which are present in a fully hybridized case, we propose a
recursive procedure for the minimization of $E_G$, i.e. the search
for the optimal set of probabilities. In spirit close to the
impurity model of the DMFT approach, we first consider that there
is only one correlated site, say $i=0$. This site is supposed to
be embedded in a reference fixed medium where all $q$'s other than
the considered site '0', are equal to 1 at beginning or to their
previous values in the self-consistent process that has to be
performed afterwards. Then the set of equations (\ref{derivp}) for
all configurations $\Gamma_0$ of site $0$ can be rewritten as

\begin{equation}
\label{derivp2} \frac{\partial E_G}{\partial
p_0(\Gamma)}=\sum_{\alpha\sigma}2e_{i=0 \alpha \sigma}
\frac{\partial ln(\sqrt{q_{0 \alpha \sigma}})}{\partial
p_0(\Gamma)}+ U_{\Gamma}
\end{equation}

The partial kinetic energies $e_{i=0 \alpha \sigma}$ of orbital
$\alpha\sigma$ at site $i=0$, generalizing (\ref{partial-kin}),
are obtained from partial projected DOS, available from any
electronic structure code, and computed with site $i=0$ embedded
in the reference medium. To ensure numerical stability when
solving system of eqs. (\ref{derivp2}) and according to the spirit
of Landau theory of Fermi liquids, the interactions are
progressively switched on from zero to their final values starting
from the probabilities of uncorrelated case. After solution, the
probabilities $p_0(\Gamma)$ are used to compute the local
$q$-factors of site $0$. If all sites are equivalent, one would
get the same results on other sites. Accordingly, the $q$-factors
of other sites are all set equal to the '0'-th ones (one would
have to repeat this impurity-like calculation if there are
inequivalent sites, i.e. crystal structures with more than one
atom per cell or disordered systems). Changing the $q$-factors
affects the partial kinetic energies $e_{i=0 \alpha \sigma}$ and
also the occupation of orbitals, as the reference medium now has
new effective hoppings. Process must be iterated till convergence.
The advantage of this way of solving iteratively eqs.
(\ref{derivp}) is that the only required ingredients to get the
solutions are partial (local) kinetic energies and occupations of
orbitals at site 0 directly obtained from partial DOS's. The price
to be paid is a greater number of electronic structure paths. It
can be easily implemented in existing codes, without searching to
get cross-terms, reducing the numerical effort to adapt our method
in these codes.

Finally, as in the one-band case, it is necessary to find the best
Slater determinant leading to optimized effective levels. One can
easily show that expression (\ref{renoG}) can be generalized for
orbital degeneracy:

\begin{equation}
\label{renoGgeneral}
\epsilon_{i\alpha \sigma}=\epsilon^0_{i\alpha
\sigma}+2e_{i\alpha \sigma}\frac{\partial ln(\sqrt{q_{i\alpha
\sigma}})}{\partial n_{i\alpha \sigma}}
\end{equation}

Again, as in the one-band case, it is necessary to perform
self-consistent calculations (for a given previously converged
$\{p_0(\Gamma)\}$ set)  till the overall convergence is reached
i.e. the on-site effective levels as well as the hoppings are
converged. Once achieved, the effective Hamiltonian $H_{eff}$ can
be used to a quasiparticles description of the system as proposed
by Vollhardt \cite{Vollhardt84}. In fact, he has shown that the
Gutzwiller method is a natural frame to obtain the parameters of
the phenomenological Landau theory of Fermi liquids. One can also
expect finite temperature extension of our method in analogy with
\cite{Gebhard91} and references therein.

\subsection{Ab-initio approach: alternative to the LDA+$U$ method}

We present now how to implement such an approach in an
\textit{ab-initio} calculation of solids. The linearized
muffin-tin orbital in the atomic sphere approximation (LMTO-ASA)
is widely used and peculiarly suited for our purpose as the basis
set has a local representation. Other \textit{ab-initio}
approaches could be used, and if they have not this local
property, one could transform the basis into a Wannier
representation. The LMTO method is well described elsewhere
\cite{Andersen75, Andersen85} and we would like to remind here
only the main results which are usefull for this paper. In the
frame of DFT-LDA band structure calculations, the LMTO method is
based on some approximations. The space is divided in atomic
spheres where the potential is spherically symmetric and
interstitial region where it is flat ("Muffin Tin" potential). In
the Atomic Sphere Approximation (A.S.A.), the spheres radii are
chosen so that the total volume of the spheres equals that of the
solid. One makes a further approximation by supposing that the
kinetic energy in the interstitial region is zero (without this
non-essential assumption, Laplace equation, as used below, should
be replaced by Helmoltz equation). In this region, the
Schr\"{o}dinger equation reduces to Laplace equation having
regular and irregular solutions: $Y_{L}(\hat{r}) r^{\ell}$and
$Y_{L}(\hat{r}) r^{-\ell-1}$ respectively. Here $L = (\ell,m)$
represents the angular momentum index and $Y_{L}(\hat{r})$ the
spherical harmonics in direction $\hat{r}=(\theta,\phi)$. For the
sphere centered at site $R$ and in the momentum index $\ell$
$(\ell=0,1,2,3)$, one finds the solution $\varphi_{R \ell \nu}$ of
the radial Schroedinger equation for a given energy $E_{\nu}$,
usually taken at the center of gravity of the occupied part of the
$\ell$-band and the energy derivative of $\varphi_{R \ell \nu}$
noted $\dot{\varphi}_{R \ell \nu}$. It can be shown that the
corresponding orbitals $\varphi_{R \ell \nu}$ and
$\dot{\varphi}_{R \ell \nu}$ are orthogonal to each other and
nearly orthogonal to the core levels. It is thus possible to build
a basis set of orbitals $\chi_{R L}$ centered at sphere of site
$R$ in the following way. Outside the sphere, in the interstitial
region $\chi_{R L}$ is proportional to the irregular solution
$Y_{L}(\hat{r}) r^{-\ell-1}$ of Laplace equation and it is
augmented (i.e. substituted according to Slater terminology) in
its own sphere by a linear combination of $\varphi_{R \ell \nu}$
and $\dot{\varphi}_{R \ell \nu}$ having logarithmic derivative
$-\ell -1$ at the radius $s_R$ of the sphere so that the orbital
is continuous and derivable at the sphere boundary. In any other
sphere $R'$, the irregular solution of Laplace equation can be
expanded in term of regular solutions in that sphere:

\begin{equation}
\label{struct-devlp}
Y_{L}(\hat{r}_R)(\frac{r_R}{a})^{-\ell-1}=-\sum_{L'}
\frac{1}{2(2\ell' +1)}S^0_{R'L',R
L}Y_{L'}(\hat{r}_{R'})(\frac{r_{R'}}{a})^{\ell'}
\end{equation}

and the orbital $\chi_{R L}$ should be augmented in sphere $R'$
with the same expansion of linear combination of $\varphi_{R'
\ell' \nu}$ and $\dot{\varphi}_{R' \ell' \nu}$ having the
logarithmic derivative $\ell'$  at the radius $s_{R'}$ of sphere
$R'$. In (\ref{struct-devlp}), $a$ is a scale factor and
$S^0_{R'L',R L}$ are the so-called "structure constants" which
depend only on the crystallographic structure of the material. In
this basis set of the orbitals $\chi_{R L}$ both Hamiltonian and
Overlap matrices can be expressed in terms of $S^0_{R'L',R L}$,
and the potential parameters $\varphi_{R \ell \nu}(s_R)$,
$\dot{\varphi}_{R \ell \nu}(s_R)$ and their logarithmic
derivatives $D_{R \ell \nu}$ and $\dot{D}_{R \ell \nu}$ at sphere
boundary. Since the structure constants $S^0_{R'L',R L}$ ,
decreasing as $r^{-\ell-\ell'-1}$ with distance, are very long
ranged for s and p orbitals, it can be more convenient to change
the basis set so that the Hamiltonian can have the Tight-Binding
(TB) form or any desired properties (like the orthogonality of
overlap). It can be achieved by adding to the regular solution of
Laplace equation an amount of the irregular solution for a given
angular momentum. It is possible to choose this amount
$\bar{Q}_\ell$ so that the transformed structure constants $S$ can
be screened with a short-range dependence with the distance or so
that the orbitals of the transformed basis set are orthogonal (the
so-called TB or most localized and orthogonal representations,
respectively). With appropriate choice for $\bar{Q}_\ell$, the
transformed structure constant matrix obeys to the following
equation:

\begin{equation}
\label{struc-scrn} S=S^{0}(1-\bar{Q}_\ell S^{0})^{-1}
\end{equation}

\noindent Matrix elements fo the Hamiltonian can be written as:

\begin{equation}
\label{Ham-LMTO1}
H_{RL,R'L'}=C_{RL}\delta_{RL,R'L'}+\Delta^{1/2}_{RL}
S_{RL,R'L'}\Delta^{1/2}_{R'L'}
\end{equation}

\noindent which is limited to first order in $(E-E_\nu)$ in the TB
representation, whereas it is valid up to second order in the
orthogonal representation (and it is even possible to add third
order correction). $C_{RL}$ determines the middle of the band
"$RL$" and $\Delta_{RL}$ its width and the strength of
hybridization. These parameters are expressed in terms of the 4
potential parameters: $\varphi_{R \ell \nu}(s_R)$,
$\dot{\varphi}_{R \ell \nu}(s_R)$, $D_{R \ell \nu}$ and
$\dot{D}_{R \ell \nu}$. It should be stressed that hybridization
between bands of different angular moments is due to the matrix
elements $S_{RL,R'L'}$ which couples $RL$-states to $R'L'$ ones.
When these matrix elements are set equal to zero for $\ell \neq
\ell'$, one obtains bands having pure $\ell$ character. This
approximation was suggested in the standard (unscreened)
representation and the resulting bands were called "canonical"
bands. In that case it would be quite easy to apply single band
Gutzwiller method (one equation per $\ell$ symmetry) without the
need of the previous fully hybridized generalization. However, as
we want to treat realistic bands, we do not use the canonical
bands in this paper. We use a scalar relativistic LMTO-ASA code
neglecting spin-orbit coupling, with the so-called "combined
corrections" which correct the ASA. The density functional
formalism, before the Gutzwiller correction explained below, was
treated within the LDA with the exchange and correlation potential
of von Barth and Hedin \cite{vonBarth72}.

The Hamiltonian of valence electrons (\ref{Ham-LMTO1}), in the
so-called orthogonal representation (or in the most localized
representation, neglecting orbital overlap) can be mapped on a
tight-binding form Hamiltonian

\begin{equation}
\label{ham2} H_{LMTO}=\sum_{i\neq
j\alpha\beta\sigma}t_{i\alpha,j\beta}c_{i\alpha\sigma}^\dagger
c_{j\beta\sigma}+
\sum_{i\alpha\sigma}\epsilon_{i\alpha\sigma}n_{i\alpha\sigma}
\end{equation}

\noindent The hoppings and on-site energies are directly outputs
of the \textit{ab-initio} calculation, as explained in
Ref.\cite{Andersen85} and by identification of expression
(\ref{ham2}) with (\ref{Ham-LMTO1}), making the correspondance: $R
\rightarrow i$, $L \rightarrow \alpha$. This opens the possibility
of treating our approach from  first principle level, without any
adjustable parameters, except the interactions $U$. They could be
however also computed from constrained LDA calculations but in the
following we rather treat them as free parameters. As in
(\ref{Hgeneral}), (\ref{ham2}) describes a full $spd$ (and
possibly $f$ as in application of this method to Plutonium, see
next section) basis. The terms $\epsilon_{i\alpha\sigma}$ account
for different on-site energies for orbitals with different angular
momentum or lying on inequivalent sites with possible crystal
field splitting between orbitals of same angular momentum, but
belonging to different irreducible group representations: it is
due to the on-site contribution of $S_{RL,RL}$ that arises in the
TB or in the nearly orthogonal representation.

 In the spirit of the Anderson model, we separate electrons into
two subsystems: delocalized electrons for which the LDA is assumed
to give reasonable results and localized electrons for which it is
well known that the LDA can lead to unphysical results. To treat
these states in a better way, and to avoid double counting, we
exclude the interaction between localized electrons ($f$ or $d$)
already taken into account in an average way in the LDA-on-site
energy

\begin{equation}
\epsilon^0_{i\alpha\sigma}=\epsilon^{\mathrm{LDA}}_{i\alpha\sigma}-U(n_f-\frac{1}{2})
\end{equation}

\noindent where $U$ is a proper combination of direct and exchange
Coulomb integral giving a true one-electron Hamiltonian $H_0$.
$n_f$  is the average number of  $f$ (or $d$) electrons given by
the LDA calculation. We then re-add an interaction part
$H_{\mathrm{int}}$ " a la Hubbard" for the localized electrons and
the full Hamiltonian $H' = H_0+H_{\mathrm{int}}$ is treated within
the previously described multiband Gutzwiller approach. In fact
this starting Hamiltonian $H'$ is the same one used in the
so-called "LDA+$U$" method\cite{Anisimov97b}, the difference being
in the way the interaction part is treated. In the LDA+$U$ method,
it is treated in a mean field Hartree-Fock like approach, which
can be questionable in case of strong correlations. It is however
a suitable way of introducing an orbital-dependent potential which
is absent in DFT formalism. In our approach the correlation is
treated exactly, within the approximation of the Gutzwiller
ansatz. Note that our method also contains an orbital-depend
potential through the renormalization of levels
(\ref{renoGgeneral}). A detailed study of the involved derivative
indeed reveals an orbital filing dependence when rewriting this
formula:

\bea
\epsilon_{i\alpha\sigma}=\epsilon^0_{i\alpha\sigma}-\frac{2
e_{i\alpha\sigma}}{n_{i\alpha\sigma}(1-n_{i\alpha\sigma})}(\frac{1}{2}-n_{i\alpha\sigma})\nonumber\\
+\frac{\partial}{\partial n_{i\alpha\sigma}} ln(\sum_{L'_i}
\sqrt{p(i\alpha\sigma: unocc,L'_i)p(i\alpha\sigma: uocc,L'_i)}
\eea

\noindent Similarly to what happens in the LDA+$U$ method, one
sees the tendency of lowering for levels with occupation greater
than one half, and a rising upwards for the less than half filled
ones (the partial kinetic energy $e_{i \alpha \sigma}$ is always
negative, it would be zero for a filled band). The difference with
LDA+$U$ method is the partial kinetic energy prefactor (instead of
$U$) and other terms that come from the derivative of the empty
and single occupied configurations.

The starting Hamiltonian $H'$ has been also used to make a link
between \textit{ab-initio} LMTO band structure calculation and a
DMFT treatment of correlations for the studies of
LaTiO$_3$\cite{Anisimov97a} and Plutonium\cite{Savrasov01}. This
last approach, assuming infinite dimension, goes beyond our
approach. We only expect to be able to describe the coherent part
of the spectrum, whereas the incoherent part leading to lower and
upper Hubbard subbands are not accessible in our model, however as
already stressed, variationally based.

The practical scheme we proposed to perform our \textit{ab-initio}
Gutzwiller approach is the following one. First, we perform a LDA
\textit{ab-initio} LMTO-ASA calculation of the solid in a given
crystal structure. This calculation provides the core and the
valence (band) electrons contribution total energy, as well as
occupations and partial kinetic energies for valence orbitals.
From these ingredients, and for a given model interaction
Hamiltonian, it is possible to evaluate the variational Gutzwiller
energy, which will be minimized, providing an optimized set of
variational configurational probabilities. Then the on-site levels
are varied according to the prescription renormalization of levels
(\ref{renoGgeneral}) as well as the adjunction of $q$-factors
(\ref{qgeneral}) which modified hoppings. New partial kinetic
energies and occupations are recalculated from the modified
Hamiltonian $H'$, until self-consistency is achieved. At the end
of procedure, the total energy, sum of the core and band energies,
is calculated, leading to the properties of the ground state. One
can then change, for example, the volume and redo the whole loop
to obtained the equilibrium properties and the most favorable
atomic configurations in the solid. Close to the Fermi level we
can also obtain an approximate \textit{ab-initio} description of
quasiparticles spectrum which enables comparison with spectroscopy
experiments for moderately correlated electron systems. That way,
one has an \textit{ab-initio} method which is
multi-configurational and variational.

In this first application, we made a slight simplification, with
respect to the general process described in last paragraph, for
the band calculation part: starting from converged LMTO potential
parameters, we build up a first order Hamiltonian in TB
representation with neglect of overlap matrix (i.e. equal to
unity) as explained in \cite{Vargas92} and references therein. As
our scheme only reorganizes valence (band) electrons, we make a
frequently used frozen core approximation assuming that the core
energy remains unaffected by this reorganization and we will now
concentrate on the band energies. It is well-known that this first
order Hamiltonian is accurate close to $E_\nu$, i.e. close to the
center of gravity of the occupied part of the bands. Far from it,
it has the effect of a slight reduction of the bandwidth, but we
have verified that it has a negligible effect on integrated
quantities: for example, before the Gutzwiller process is switched
on, we have checked that the band energies calculated from the
third order Hamiltonian and from the first order one with the
recursion process described below, are in excellent agreement. We
used a full $spdf$ basis set with hoppings up to second nearest
neighbors. For all 7 inequivalent orbitals, in cubic environment
from the overall 16 orbitals, we performed a real space recursion
procedure \cite{Haydock84} to get the partial projected densities
of states (DOS) from which all needed quantities, like occupancies
or band energies, can be calculated. These partial DOS are
obtained from the imaginary part of diagonal elements of a Green
function, which are developed in a continued fraction expansion up
to a given level. This level is chosen so that a convergence
criterium is reached, i.e. adding one more level does not affect
the result. Practically we took 40 steps of recursion. Various
terminators (the well-known square root terminator, or more
elaborated ones in presence of gaps \cite{TurchiDucastelle}) are
then used to close the continued fraction expansion. A full
self-consistent approach within the Gutzwiller loop, using third
order Hamiltonians and including spin-orbit coupling, is still in
preparation, and some intermediary results will be given in next
section.

\section{Application to Plutonium}
We now give a simple application of the present method to
Plutonium which is a good test case. Pu lies between light
actinides with itinerant 5$f$ electrons and heavy actinides with
localized 5$f$ electrons. The competition between these two
electronic regimes in Pu is responsible for a lot of unusual
properties as large values of the linear term in the specific heat
coefficient and of the electrical resistivity or a very complex
phase diagram. The ground state $\alpha$ phase (monoclinic with 16
atoms by cell) is known to be well described by \textit{ab-initio}
DFT-LDA calculations, whereas for the high temperature $\delta$
phase (fcc), the calculated equilibrium volume is of the order of
30 percent smaller than the experimental one. It is very important
to reproduce the properties of Pu to take into account the
delicate balance between the itinerancy of the $f$ electrons and
the large intra atomic Coulomb interaction. This requires a much
more complicated theory for the electrons than the LDA which is
like a mean-field treatment of the correlations
\cite{Lundqvist83}. Recently several attempts to go beyond LDA
have given a new understanding of the $\alpha$-$\delta$
transition. In the LDA+$U$ method \cite{Bouchet00} an
orbital-dependent correction, treated in the mean-field
approximation is added to the LDA functional. These calculations
have showed how the equilibrium volume is improved in comparison
to previous results using LDA, and how an augmentation of the
orbital moment is observed following Hund's rules, reducing the
total magnetic moment in agreement with experiments. Going a step
beyond the LDA+$U$, Savrasov \textit{et al} \cite{Savrasov00} have
used an implementation of DMFT. The LDA+$U$ can be viewed as the
static approximation of the DMFT. With this dynamical treatment of
the $f$-electrons they have recovered the experimental equilibrium
volume of $\delta$-Pu, the photoemission peak at the Fermi level
and given an understanding picture of the transition between
$\alpha$ and $\delta$ phases. Different approaches, using the
spin-polarized generalized gradient approximation (GGA) and
antiferromagnetic configurations \cite{Wang00, Kupetov03,
Soderlind01} have well reproduced the ground state properties of
$\delta$-Pu. All these works show how a spin/orbital polarization
is crucial to describe the $\delta$-phase. In the Gutzwiller
method, the correlations, via the $q$-factors, are supposed to
reduce the hoppings, and so to weaken the covalency character of
the bonding, and consequently the attraction between atoms. Thus
we expect to increase the interatomic distance, leading to a
greater equilibrium volume. Of course, the same approach has to be
performed for $\alpha$ and $\delta$ phase.

An extra difficulty arises from the Atomic Sphere Approximation
(ASA) of the LMTO method: the atomic potential, inside an atomic
"muffin-tin" sphere, is spherized, or equivalently, the true
"full" potential is approximated by its first $\ell=0$ component.
This approximation greatly simplifies the calculation, as the wave
function basis in a sphere, used to build the LMTO set, can be
factorized in a product of a radial wave function and a spherical
harmonics  as explained above. It presents however the
shortcomings that, it is not a "full potential" approach and
forbids to change the symmetry when making comparison between
structures. We overcome this difficulty here by performing the
calculation in a fcc structure browsing different volume: it is
correct for the $\delta$ phase, but the $\alpha$ phase will be
replaced by a "pseudo"-$\alpha$  phase, in a fcc structure, having
however the same density than the experimental one.

The valence states taken into account in the LMTO part were the
7$s$, 6$p$, 6$d$, 5$f$ of Pu with 16 fully hybridized orbitals per
site, the remaining orbitals being treated as core states. In this
first approach, as we concentrate more on the correlation effects,
we neglect, however important for this heavy element, the
spin-orbit coupling. The crystal field splitting on $f$ orbitals
(and other ones), is directly accounted by the LMTO method,
lifting the $f$ degeneracy in the 6-fold (including spin) $T_1$,
the 6-fold $T_2$ and 2-fold $A_2$ symmetries. Finally, the
interaction $H_{int}$, added to $H_0$, is simply given by the same
local term between electrons on different $f$-orbitals

\begin{equation}
H_{\mathrm{int}}=\frac{U}{2}\sum_{i,\alpha\sigma\neq \beta
\sigma'}n_{i\alpha\sigma}n_{i\beta\sigma'}
\end{equation}

\noindent neglecting any exchange term as done in
\cite{Bouchet00,Savrasov00}. In this simplified paramagnetic
version, the number of inequivalent atomic configurations,
necessary to perform the Gutzwiller part, reduces to 14 because
all atomic configurations having the same electronic occupancy are
equivalent in this model. Similarly, we took an average occupation
per $f$ orbital in the expression of $q$-factors, leading to a
single $q$ for all $f$ orbitals, regardless to crystal field
splitting. It was, however, included for the on-site levels
renormalization, since the partial kinetic energy and occupations
are not exactly equal for different symmetries. We have
nevertheless checked this assumption by performing a much heavy
calculation, including 3 different $q$'s, one per crystal symmetry
with 7x7x3 = 147 variational parameters: the final result was not
sensitive to this detail. It reflects the small $f$-crystal field
splitting in Plutonium, producing very similar occupations and
partial kinetic energies.

The Coulomb interaction $U$ could be also provided by constrained
LDA calculations. In that sense, it would not be an adjustable
parameter. However, we did not recalculate its value and took it
from literature, close to 0.3Ry, as in the LDA+DMFT calculation of
Savrasov et al. \cite{Savrasov01}, or as in the LDA+$U$
calculation of Bouchet et al. \cite{Bouchet00}. An improved
version of calculation, including exchange interaction, as in the
degenerate Hubbard model, with one $q$-factor per symmetry, will
be used in a forthcoming paper, in which we will investigate also
ferromagnetic and antiferromagnetic ground states. In this work we
just want to appreciate the effect of our method and of the
Gutzwiller approximation on a simple case,where there exists known
results with other methods.

\begin{figure}
\begin{center}
\rotatebox{270}{\includegraphics[scale=0.3]{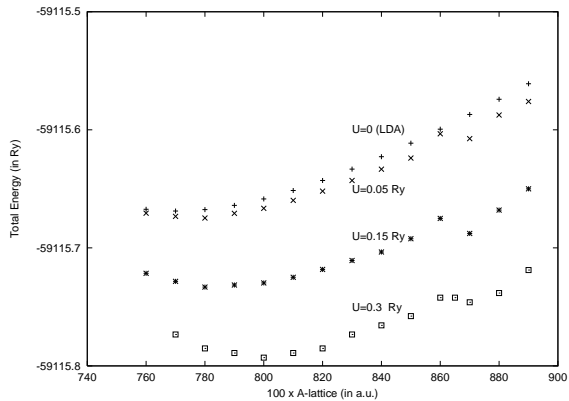}}
\caption{
Total energy of fcc-Pu versus volume for different values of the
interaction $U$. $U=0$ corresponds to a LDA calculation.}
\label{etotu} \end{center}
\end{figure}

The total energy versus volume for fcc-Pu and different values of
the interaction $U$ is presented in Fig. \ref{etotu}. The curve
$U=0$ corresponds to a LDA calculation. As previously found in
several works the minimum of this curve is very low ($\sim 7.70$
ua ) compared to the experimental value of the $\delta$ phase
(8.60 ua) and closer to the $\alpha$ phase value (8.0 ua). In fact
there is no sign of the correlated $\delta$ phase in the $U=0$
calculation. As we turn on the correlations, a new feature appears
in the curves, almost instantly. We observe a new energy minimum
close to the experimental volume of the $\delta$ phase. Moreover
the first minimum increases to approach the value of the
experimental $\alpha$ volume, showing that correlations are
already important to reproduce the properties of this phase. For a
value of $U$ close to 0.3 Ry, the two minimums correspond to the
experimental values of $\alpha$ and $\delta$-Pu. This double-well
feature of the total energy curve of Pu was previously discovered
by Savrasov \textit{et al}\cite{Savrasov01}, using a DMFT
approach. In our calculations the first minimum is the lower one,
since the $\alpha$ phase is the ground state for Pu, and we
haven't added any temperature effect in our calculations. As $U$
increases we see a tendency of the two minimums to be closer. In
fact the energies of the two phases are very similar and a small
perturbation, for example the temperature, can be sufficient for
the phase transition. Of course the model studied in this work is
still very simple and we don't want to conclude too far but we
think that it already contains the key ingredients (competition
between localization and delocalization, atom-like or bands-like
descriptions) to reproduce the main characteristics of Plutonium
phase diagram. Due to the roughness of our first approach, the
(rather) good agreement for the equilibrium properties, may be
incidental or due to some compensation effect, and the
disagreement with other aspects (like bulk modulus, see below) is
not surprising. Indeed, it is well known the the spin-orbit
coupling is a key ingredient for this element: the splitting
between 5/2 and 7/2 states could give significant differences in
occupation and kinetic energies. One may expect then a difference
between $q_{5/2}$ and $q_{7/2}$, and obtain localized and less
localized behaviors as suggested by P\'enicaud \cite{Penicaud97}
who proposed to split $f$ states between localized and more
delocalized ones to explain the properties of Plutonium. The
freezing of $f$-states to similar occupation in our present
calculation could be responsible for the high value of the bulk
modulus (637 GPa) we get, in contrast with the experimental value
of 30 GPa \cite{Ledbetter76}. Primary result with an improved
version involving third order LMTO Hamiltonian full
self-consistent computation, neglecting yet spin-orbit coupling,
reduces this value to 196 GPa, which is slightly better than the
LDA result of 214 GPa \cite{Soderlind94}.

This \textit{ab-initio} Gutzwiller approach is able to handle
correctly the correlation aspects without loosing the
\textit{ab-initio} adjustable parameters free aspect of the more
familiar DFT-LDA, and that way, corrects the deficiency of this
method. It gives similar results to the methods that account for
many-body effects like the LDA+DMFT of Ref.\cite{Savrasov01} from
the \textit{ab-initio} levels or that can have an orbital
dependent potential like in the LDA+$U$ calculation of
Ref.\cite{Bouchet00}, which is impossible to DFT-LDA approach. On
another hand, we stress again that our approach is clearly
variational, and is able to provide an approximate ground state in
contrast with those of Refs. \cite{Savrasov01} and
\cite{Bouchet00}.

\begin{figure}
\begin{center}
\rotatebox{270}{\includegraphics[scale=0.3]{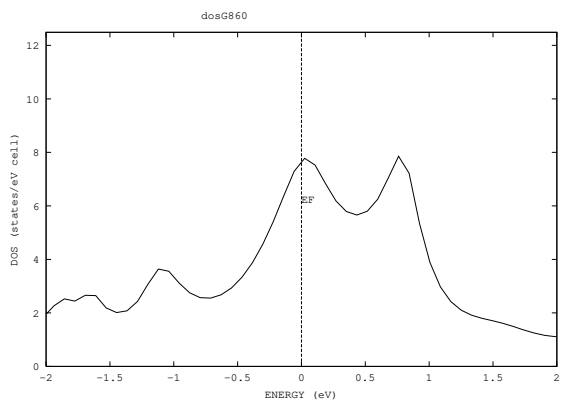}}
\caption{
Quasiparticles density of states obtained from Gutzwiller method
for Plutonium in $\delta$ phase.} \label{QP-DOS}
\end{center}
\end{figure}

The effective optimized Hamiltonian $H'$, was used to compute
quasiparticles density of states, in the vicinity of Fermi energy.
The result, shown on Fig. \ref{QP-DOS}, is restricted to an energy
window of 2eV on both sides of Fermi level: further, the spectrum
would be in the region of the Hubbard subbands, which are beyond
the scope of our approach (it should be necessary to include
fluctuations to get the incoherent part of  the spectrum). It is
to be stressed that our result compares well, in the presented
region, with the more elaborated LDA+DMFT result of
Ref.\cite{Savrasov01}. Both results are also in good agreement
with the photoemission experiments of Arko et al. \cite{Arko00}.
The peak at the Fermi level, which results mainly from the
reduction of hoppings due to $q$-factors, associated with a small
shift of Fermi level due to the renormalization of levels
(\ref{renoGgeneral}), has the consequence of a significant
improvement of the electronic specific heat contribution,
multiplied by a factor of 2 with respect to the LDA value. Our
result, of the order of 13 mJ K$^{-2}$ mol$^{-1}$is however yet
far from the experimental value of 64 mJ K$^{-2}$ mol$^{-1}$ found
by Lashley \textit{et al}\cite{Lashley03}. The $q$-factor,
responsible for this increase of the DOS at the Fermi level, is of
the order of 0.8 in the equilibrium $\delta$ phase. This moderate
renormalization is due to hybridization of the $f$ states with the
low lying $p$-states which lead to a significant partial $f$
kinetic energy, greater than what it would be considering only the
narrow group of predominant $f$-character states. With a naive
single $f$-band argument, if we had used for example canonical
(i.e. unhybridized) bands, one obtains a much reduced $q$-factor
close to .3 \cite{BouchetThese}. It can be shown that the
$q$-factor is the spectral weight of the quasiparticles. A
moderate value, as obtained by our realistic calculation (i.e.
fully hybridized bands), means a rather high weight: for
independent particles it would be equal to one. It validates a
quasiparticles picture description, allowing \textit{a posteriori}
comparison we did with spectroscopy experiments. At the volume of
the "pseudo" $\alpha$ phase, this $q$ factor reduces to .9,
indicating that the electrons are less correlated in this phase,
which can explain the relative success of its description by LDA
calculation.

\section{Conclusion}
To conclude, we have generalized the density matrix approach to
Gutzwiller method for the degenerate Hubbard Hamiltonian. We have
shown that we can express the total energy in the Gutzwiller state
in terms of the different probabilities of configurations.
Moreover to apply the method to cases of physical interest we have
developed this method for inequivalent sites and for different
orbital symmetries. In this way we have given the expression of
the different $q$ factors which renormalize the hopping terms and
an expression to renormalize the on-site energies in the
Gutzwiller context. This method is limited to ground state
properties but can extend to finite temperature and low-frequency
excitations in analogy with the work of Gebhard \cite{Gebhard91}.
Of course, as a quasiparticle approach, this method is limited to
cases where the Fermi-liquid theory is valid, i.e. close to Fermi
energy. Thereafter we have have described a simple implementation
of our method in a \textit{ab-initio} calculation as the LMTO
method. To give an example, we have applied this technique to the
particular case of Pu in fcc structure. In despite of the
simplicity of our model, we were able to extract interesting
results such as the double-well feature in the energy-volume curve
and more generally improve the LDA results. Our results compare
well with previous works.

\section{Acknowledgments}
This work was performed under the auspices of the Commissariat \`a
l'Energie Atomique under Contract No 9M 1898. The authors wish to
thank R.C Albers of Los alamos National Laboratory for supporting
of the present work. We are also grateful to F. Jollet, A.
Pasturel and A. Georges for stimulating discussions.


\begin{thebibliography}{99}
\bibitem{FuldeBook}
P. Fulde, \textit{Electron Correlations in Molecules and Solids},
vol. 100 ( Solid-State Sciences, Springer, 1995).

\bibitem{Hohenberg64}
P. Hohenberg and W. Kohn, Phys. Rev., \textbf{136}, 864 (1964).

\bibitem{Kohn65} W. Kohn and L. J. Sham, Phys. Rev.,
\textbf{140}, 1133(1965).

\bibitem{Mattuck}
R.D. Mattuck, \textit{A Guide to Feynman Diagrams in the Many-Body
Problem}, second edition (Dover, 1974).

\bibitem{Kotliar86}
 G. Kotliar and A.R. Ruckenstein, Phys. Rev. Lett., \textbf{57},
 1362 (1986).

\bibitem{Mori}
H. Mori, Prog. Theor. Phys., \textbf{33}, 423 (1965).

R. Zwanzig, \textit{Lectures in Theoretical Physics}, Vol. 3
(Interscience, New York, 1961).



\bibitem{Georges96}
 A. Georges, G. Kotliar, W. Krauth and J. Rozenberg, Rev. Mod.
 Phys., \textbf{68}, 13 (1996).

\bibitem{Anisimov97b}
V. I. Anisimov,  F. Aryasetiawan and A. I. Lichtenstein, J. Phys.:
Condens. Matter, \textbf{\textbf{9}}, 767 (1997).

\bibitem{Anisimov97a} V. I. Anisimov, A. I. Poteryaev, M. A.
Korotin, A. O. Anokhin and G. Kotliar, J. Phys.: Condens. Matter,
\textbf{9}, 7359 (1997).

\bibitem{Savrasov01} S. Y. Savrasov, G. Kotliar and E.
Abrahams, Nature (London), \textbf{410},793 (2001).

\bibitem{Gutzwiller63}
M. C. Gutzwiller, Phys. Rev. Lett., \textbf{10}, 159 (1963).

\bibitem{Gutzwiller65}
M. C. Gutzwiller, Phys. Rev., \textbf{137}, A1726 (1965).


\bibitem{Hubbard64} J. Hubbard, Proc. Roy. Soc. London,
\textbf{A 276}, 238 (1963).

\bibitem{bookGebhard}
F. Gebhard, \textit{The Mott Metal-Insulator Transition- Models
and Methods}, Tracts in Modern Physics, \textbf{137}, Springer
(1997).


\bibitem{Vollhardt84}
D. Vollhardt, Rev. Mod. Phys., \textbf{56}, 99 (1984).


\bibitem{Nozieres}
P. Nozi\`{e}res, \textit{Magn\'etisme et localisation dans les
liquides de Fermi}, Cours du Coll\`{e}ge de France, Paris (1986).


\bibitem{BrinkmannRice70}
W.F. Brinkmann and T.M. Rice, Phys. Rev. B, \textbf{2}, 1324
(1970).

\bibitem{Mayou88}
D. Mayou, D. N. Manh and J.-P. Julien, Solid State Comm.,
\textbf{68}, 665 (1988).

\bibitem{Julien99}
J.-P. Julien, Physica B, \textbf{259}, 757 (1999).

\bibitem{Julien05}
J-P. Julien and J. Bouchet, Physica B, \textbf{359-361}, 783
(2005).

\bibitem{Bunemann98b}
J. B\"{u}nemann, W. Weber and F. Gebhard, Phys. Rev. B,
\textbf{57}, 6806 (1998).

\bibitem{BGW-JPhConMatter97}
J. B\"{u}nemann, F. Gebhard and W. Weber, J. Phys.: Condens.
Matter, \textbf{9}, 7343 (1997).

\bibitem{MetznerVollhardt87}
W. Metzner and D. Vollhardt, Phys. Rev. Lett., \textbf{59}, 121
(1987).

\bibitem{Sugano} S. Sugano, Y. Tanabe and H. Kamimura, Pure Appl.
Phys., \textbf{33}, 38 (1970).

\bibitem{Bunemann97c}
J. B\"{u}nemann and W. Weber, Phys. Rev. B, \textbf{55}, 4011
(1997).

\bibitem{Okabe}
T. Okabe, J. Phys. Soc. Jpn., \textbf{66}, 2129 (1997).

\bibitem{Hasegawa97b}
H. Hasegawa, J. Phys. Soc. Jpn., \textbf{66}, 1391 (1997).

\bibitem{Lu}
J.P. Lu,  Internat. Journal of Modern Phys., \textbf{B10}, 3717
(1996).

\bibitem{Andersen75} O.K. Andersen,
Phys. Rev. B, \textbf{12}, 3060 (1975).

\bibitem{Andersen85}O. K. Andersen, O. Jepsen and D. Gloetzel,
\textit{Highlights of Condensed Matter Theory - Varenna notes -
Proceedings of The International School of Physics Enrico Fermi},
North Holland, New York(1985).

\bibitem{vonBarth72}
U. von Barth and L. Hedin, J. Phys. C, \textbf{5}, 1629 (1972).

\bibitem{Vargas92}
P. Vargas, \textit{Methods of Electronic Structure Calculations},
147 (Proceedings of the Miniworkshop on Methods on Electronic
Structure Calculations, Trieste, 1992).

\bibitem{Haydock84}
R. Haydock, in \textit{Solid State Physics}, \textbf{35}, 215
(Academic Press, New York,1980).

\bibitem{TurchiDucastelle}
P. Turchi, F. Ducastelle and G. Tr\'{e}glia, J. Phys. C,
\textbf{15}, 2891 (1982).

32\bibitem{Lundqvist83} S. Lundqvist and N.H. March,
\textit{Theory of the inhomogeneous electron gas} (Plenum Press,
New York, 1983).

\bibitem{Bouchet00}
J. Bouchet, S. Siberchicot,  A. Pasturel
and F. Jollet, J. Phys.: Condens. Matter, \textbf{12}, 1723
(2000).

\bibitem{Savrasov00}
S. Y. Savrasov and G. Kotliar, Phys. Rev. Lett., \textbf{84}, 3670
(2000).

\bibitem{Wang00}
Y. Wang and Y. Sun, J. Phys.: Condens. Matter, \textbf{12}, L311
(2000).

\bibitem{Kupetov03}
A.L. Kupetov and S.G. Kupetova, J. Phys.: Condens. Matter,
\textbf{15}, 2607 (2003).

\bibitem{Soderlind01}
P. S\"{o}derlind, Europhys. Lett., \textbf{55}, 525 (2001).

\bibitem{Penicaud97}
M. P\'{e}nicaud, J. Phys.: Condens. Matter, \textbf{9}, 6341
(1997).

\bibitem{Ledbetter76}
H.M. Ledbetter and R.L. Moment, Acta Metallurgica, \textbf{24},
891 (1976).

\bibitem{Soderlind94}
P. S\"{o}derlind, \textit{Theoretical Studies of Elastic, Thermal
and Structural Properties of Metals} (PhD Thesis, 1994).

\bibitem{Arko00}
A.J. Arko, J.J. Joyce, L. Morales, J. Wills, J.C. Lashley, F.
Wastin and J. Rebizant, Phys. Rev. B, \textbf{62}, 1773 (2000).

\bibitem{Lashley03}
J.C. Lashley, J. Singleton, A. Migliori, J.B. Betts, R.A. Fisher,
J.L. Smith and R.J. McQueeney, Phys. Rev. Lett., \textbf{91},
205901 (2003).

\bibitem{BouchetThese}
J. Bouchet, \textit{Etude du Plutonium en Phase $\delta$ et de ses
Alliages avec les El\'{e}ments de la Colonne IIIB} (PhD Thesis,
unpublished, Universit\'{e} Paris VI, 2000).

\bibitem{Gebhard91} F. Gebhard, Phys. Rev. B, \textbf{44}, 992
(1991).

\end{thebibliography}
\end{document}